\documentclass[showpacs,pra,aps,preprint,onecolumn,superscriptaddress]{revtex4-1}
\usepackage{color}
\usepackage{dcolumn}
\usepackage{graphicx,amsmath,amsfonts,amssymb}
\usepackage{tikz,subfigure,graphicx}
\usetikzlibrary{snakes}
\newcommand{\eps}{\varepsilon}
\newcommand{\sinc}{\mathrm{sinc}}
\renewcommand{\vec}[1]{\mathbf{#1}}

\begin{document}
	\title{Effects of imperfect angular adjustment on plasmonic force}
	
\author{Leonid Frumin}
\affiliation{Institute of Automation and Electrometry,  Russian Academy of Sciences, Siberian Branch, 1 Koptjug Avenue, Novosibirsk 630090, Russia}
\affiliation{Novosibirsk State University, 2 Pirogov Street, Novosibirsk 630090, Russia}
\author{Alexander Tusnin}
\affiliation{Institute of Automation and Electrometry, Russian Academy of Sciences, Siberian Branch, 1 Koptjug Avenue, Novosibirsk 630090, Russia}
\affiliation{Novosibirsk State University, 2 Pirogov Street, Novosibirsk 630090, Russia}
\author{Oleg Belai}
\affiliation{Institute of Automation and Electrometry,  Russian Academy of Sciences, Siberian Branch,  1 Koptjug Avenue, Novosibirsk 630090, Russia}
\author{David Shapiro}
\affiliation{Institute of Automation and Electrometry,  Russian Academy of Sciences, Siberian Branch, 1 Koptjug Avenue, Novosibirsk 630090, Russia}
\affiliation{Novosibirsk State University, 2 Pirogov Street, Novosibirsk 630090, Russia}

\begin{abstract}
The attractive plasmon force between two metallic walls when the electromagnetic wave propagates through a narrow slit has been studied earlier for parallel plates and normal incidence. In present paper the effects of imperfect adjustment of plates and laser beam are analyzed. The change of force for non-parallel plates is shown to be of the first order in inclination angle when the wedge is along wave propagation and of the second order for transverse case. The small incidence angle leads to decrease in force due to the antisymmetric waveguide mode appearing in the slit.
	
\end{abstract}

\maketitle
\section{Introduction}

Since the 1980s, Micro-Electro-Mechanical Systems (MEMS) have rapidly developed and become an essential part of the technology for micro-devises \cite{Gardner:2001,969936,rebeiz2004rf}. This industry produces controllers, accelerometers, gyroscopes having various applications, from common cell phones \cite{ko2007trends} to biosensors \cite{grayson2004biomems}. Today a new type
of fast micro-controllers appears as a merge of MEMS and micro-optics: Micro-Opto-Electro-Mechanical Systems (MOEMS).
Being based on the interaction between electromagnetic field and solid body MOEMS provides new possibilities in the nano- and micro-particles manipulation.

In the case of metallic particle the surface plasmon polaritons are excited at the interface metal-dielectric, where the real part of dielectric function change its sign, and cause the plasmon force \cite{Arias-Gonzalez:03}.
The forces, with collective plasmon resonances excited by the laser field, are  studied experimentally for the dielectric sphere near a conducting plate \cite{volpe2006surface} and theoretically between two close metallic nanospheres \cite{chu2007laser}. They are challenging for optical trapping and laser tweezers \cite{juan2011plasmon,jones2015optical}. In particular case of slit between plane-parallel metallic plates the attractive plasmon force has been predicted \cite{0295-5075-94-6-64002,0957-0233-22-9-094008}
with properties determined by the geometry, conductivity, and the light polarization. Magnitude of this force for gold walls and normal incidence is of the order of nanonewtons, hence it becomes an important experimental issue. To study this effect closely an experiment is carried out with the Nano Force Facility \cite{nies2016experimental,0026-1394-53-4-1031}. At the same time previous theoretical studies did not consider the possible experimental uncertainties such as  misalignment of the plates or laser beam.

The goal of present paper is to calculate corrections to the force at small deviation from the plane-parallel geometry. In the first order the perturbation is a sum of corrections over different uncertainties, and then the additions can be estimated separately. In Sec.~\ref{sec:plates} the wedge  parallel (A) and perpendicular (B) to the propagation is studied. The oblique incidence is treated in Sec.~\ref{sec:oblique}. The field (A) and Maxwell's tension (B) are calculated within the lower modes approximation.
Sec.~\ref{sec:conclusions} summarizes the results.

\section{Non-parallel plates}\label{sec:plates}
\begin{figure}
	\begin{tikzpicture}[scale=1.25]
	\draw [very thick,red,->] (-0.3,-1)--(-0.1,-0.4);
	\path[draw,fill=yellow]  {[snake=zigzag](-2.5,0)--(-2.5,3) --(-0.5,3)}--(-0.5,0)--(-2.5,0);
	\path[draw,fill=yellow]  {[snake=zigzag](2.5,0)--(2.5,3) --(0.5,3)}--(0.5,0)--(2.5,0);
	\draw [blue,thick,->] (-2.7,0)--(2.7,0);
	\draw [black] (0.5,-0.25) node {\large$\ell$};
	\draw [black] (-0.7,-0.25) node {\large$-\ell$};
	\draw [blue] (2.6,-0.25) node {\large$x$};
	\draw [blue,thick,->] (0,-1)--(0,3.1);
	\draw [blue] (-0.25,3) node {\large$z$};
	\draw [fill=white,thick,opacity=1] (0,0) circle (5pt);
	\draw [] (-0.12,-0.12)--(0.12,0.12);
	\draw [] (0.12,-0.12)--(-0.12,0.12);
	\draw [blue] (0.25,0.25) node {\large$y$};
	\draw (-0.5,-1.25) node {\large$\vec{k}_0$};
	\draw (-1.9,-0.7) node {\large$\vec{E} = (E_x,0,E_z)$};
	\draw (1.7,-0.7) node {\large$\vec{H} = (0,H_y,0)$};
	\draw (0,-0.8) arc (-90:-103:1);
	\node[] at (263:1.2)  {\large$\gamma$};
	\end{tikzpicture}
	\caption{Geometric scheme of the slit with $p$-wave. At $\gamma=0$ the  incidence is normal.
	Here $\vec{E},\vec{H}$ are vectors of electric and magnetic field, $\vec{k}_0$ is the wave vector.}
	\label{fig:scheme}
\end{figure}
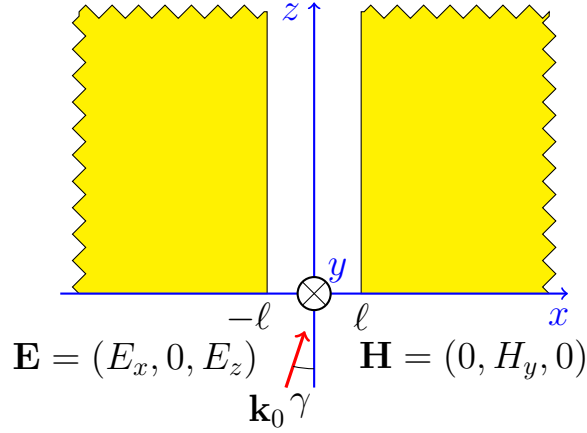

The light-induced force had been studied theoretically for normal incidence in plane-parallel geometry \cite{0295-5075-94-6-64002,0957-0233-22-9-094008}. The geometrical scheme is shown in Fig.~\ref{fig:scheme}.  The plane wave has $p$-polarization with magnetic field vector along axis $y$, since only this state excites surface plasmons. The parallel metal plates at the distance $2\ell$ are considered as infinite in $x,y,z$ directions. The wavelength of radiation $\lambda$ is assumed much greater than the half-width $\ell$. In the opposite case waveguide modes are excited, and then decrease the amplitude $h_0$ of field at the slit entrance \cite{Shapiro16}. Plasmons are generated at the surface of metal with dielectric permittivity $\eps_M=\eps_1+i\eps_2$, where $\eps_1<0,|\eps_1|\gg\eps_2\sim1$.

Remind the main relations for a perfect plane-parallel slit. Write electric and magnetic fields of monochromatic wave with frequency $\omega$ as
\[
E_i=e_i e^{-i\omega t}+\mathrm{c.c.},\quad H_i=h_i e^{-i\omega t}+\mathrm{c.c.},
\]
where $i=x,y,z$ is the Cartesian index, $\mathrm{c.c.}$ means complex conjugated terms. The Maxwell tension tensor is
\begin{eqnarray}\label{sigma0}
\sigma_{xx}=\frac{|e_x|^2-|h_y|^2-|e_z|^2}{4\pi}
\simeq\frac{h_0^2e^{-2\beta_2z}}{4\pi\sqrt{|\eps_1|}k_0l},\label{tension}
\end{eqnarray}
where $k_0=\omega/c$ is the wavenumber of incident field in free space, $c$ is the speed of light, $h_0$ is the amplitude.
The propagation constant in the slit along $z$ is $\beta=\beta_1+i\beta_2$, where
\begin{equation}\label{beta}
\beta_1\simeq k_0+\frac{1}{2l\sqrt{|\eps_1|}},\quad
\beta_2=\frac{\eps_2}{4l|\eps_1|^{3/2}}.
\end{equation}
If the plates are infinite along $z$ the unperturbed force per unit length in $y$-direction of is
\begin{equation}\label{unperturbed}
f_0\simeq\frac{\sigma_{xx}}{2\beta_2}=
\frac{h_0^2\lambda|\eps_1|}{4\pi^2\eps_2}.
\end{equation}
The value $h_0$ at the entrance of slit $(x,z)=(0,0)$ is determined by the interference of incident and reflected waves. In the limiting case  of very narrow slit $\ell/\lambda\ll1$ their amplitudes are equal and $h_0=2H_0$, where $H_0$ is the amplitude of plane incident wave $H_y(\vec{r})=H_0e^{i\vec{k}\cdot\vec{r}}$. For wider slit $h_0$ has oscillating $\ell$-dependence die to excitation of higher waveguide modes \cite{Shapiro16}.

\subsection{Longitudinal wedge}

\begin{figure}
	\subfigure[]{
	\begin{tikzpicture}[scale=3]
	\draw [line width=1.5pt,red,->,opacity=1] (0,0.4)--(0,0.7);
    \draw (-0.1,0.6) node {\large$\vec{k}_0$};
    \draw[->] (-1,0)--(1,0) node at (.95,-0.1) {\large$x$};
	\draw[->] (0,0)--(0,1) node at (.1,0.95) {\large$z$};
	\draw[line width=1.5pt,blue] (0.3,0)--(0.5,1);
	\draw[line width=1.5pt,blue] (-0.3,0)--(-0.5,1);
	\draw[dashed] (0.3,0)--(0.3,1);
	\draw[dashed] (-0.3,0)--(-0.3,1);
	\draw (0.3,0.8) arc (90:80:1) node at (0.4,0.9) {\large$\alpha$};
	\draw (0,-0.2) node {\large$0$};
			\draw (-0.15,-0.1) node {\large$y$};
	\draw [fill=white,thick,opacity=1] (0,0) circle (2.5pt);
		\draw [] (-0.06,-0.06)--(0.06,0.06);
		\draw [] (0.06,-0.06)--(-0.06,0.06);
	\end{tikzpicture}
}
\subfigure[]{
	\begin{tikzpicture}[scale=3]
	\draw[->] (-1,0)--(1,0) node at (.95,-0.1) {\large$x$};
	\draw[->] (0,0)--(0,1) node at (.1,0.95) {\large$y$};
	\draw[line width=1.5pt,blue] (0.3,0)--(0.5,1);
	\draw[line width=1.5pt,blue] (-0.3,0)--(-0.5,1);
	\draw[dashed] (0.3,0)--(0.3,1);
	\draw[dashed] (-0.3,0)--(-0.3,1);
	\draw (0.3,0.8) arc (90:80:1) node at (0.4,0.9) {\large$\alpha$};
	\draw (0,-0.2) node {\large$0$};
	\draw (-0.15,-0.1) node {\large$z$};
	\draw [fill=white,thick,opacity=1] (0,0) circle (2.5pt);
	\draw[fill=black] (0,0) circle (0.7pt);
\draw [very thick,red,fill=white,opacity=1] (0,0.55) circle (2pt);
	\draw[red,fill=red] (0,0.55) circle (0.7pt);
    \draw (-0.15,0.5) node {\large$\vec{k}_0$};
\end{tikzpicture}
}		
	\caption{Scheme of longitudinal (a) and transverse (b) wedge: the inclined metallic walls (thick solid line), parallel walls (dashed).}\label{fig:longitudinal}
\end{figure}
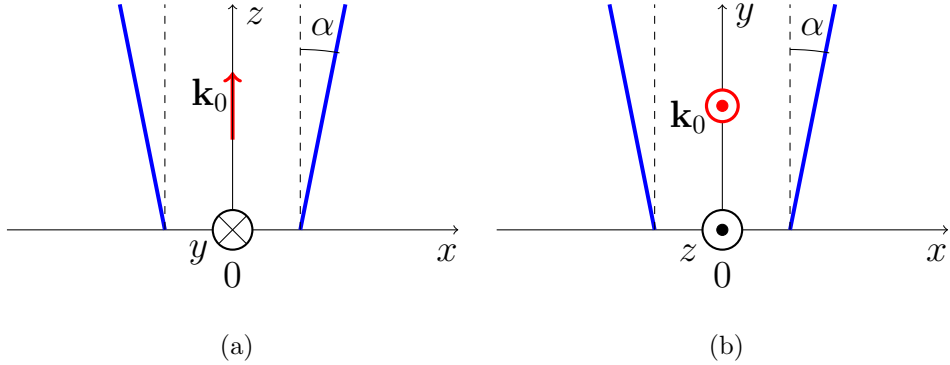

\begin{figure}\centering
	\begin{tikzpicture}[scale=1.5]
\draw[green!80!black,->] (0,2.37) arc (90:100:2.37);
\draw (-0.2,2.25) node {\large$\varphi$};
	\draw (-0.15,1.55) node {\large$r_0$};   
	\draw[thick,->] (0,0)--(0,3.5);
	\draw[dashed]  (0.405,1.69)--(0.405,0);
	\draw[dashed]  (-0.405,1.69)--(-0.405,0);
	\draw (0.43,-0.2) node {\large$\ell_0$};
	\draw (-0.48,-0.2) node {\large$-\ell_0$};
	\draw[thick,->] (-1.5,0)--(1.5,0);
	\draw (-0.15,3.4) node {\large$z$};
	\draw (1.4,0.15) node {\large$x$};
	\draw (0,1.73) arc (90:104:1.73);
	\draw (0,1.73) arc (90:76:1.73);
	\draw (0,2.15) arc (90:104:2.15);
	\draw (0,2.15) arc (90:76:2.15);
	\draw (0,2.58) arc (90:104:2.58);
	\draw (0,2.58) arc (90:76:2.58);
	\draw (0,3) arc (90:104:3);
	\draw (0,3) arc (90:76:3);
	\draw (0,0.867) arc (90:76:0.867);
	\draw (0.1,0.7) node {\large$\alpha$};
	\draw (0,0)--(0.73,3);
	\draw (0,0)--(-0.73,3);
	\draw (0,-0.25) node {\large$0$};
	\draw[line width=3pt,blue] (0.405,1.69)--(0.73,3);
	\draw[line width=3pt,blue] (-0.405,1.69)--(-0.73,3);
	\end{tikzpicture}
	\caption{Polar coordinates for the problem of vertically inclined walls: $r$ is counted from the intersection point 0 of the wall extensions, $\varphi$ from axis $z$. The slit is bounded by metal walls $\varphi=\pm\alpha$. The distance from the origin to slit entrance is $r_0=\ell_0/\alpha$, where $\ell_0$ is the slit half-width. Cartesian coordinate $z$ is counted from the entrance $(r,\varphi)=(r_0,0)$.}
	\label{f:polar}
\end{figure}
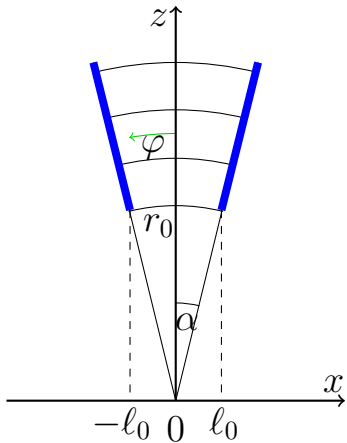

Let the walls are inclined symmetrically, as shown in Fig.~\ref{fig:longitudinal}~(a), where $\alpha\ll1$ is the wedge half-angle. The walls  can be described by a pair of linear equations
$x=\ell_0+\alpha z, x=-\ell_0-\alpha z.$ The Helmholtz equation for magnetic field in $(x,z)$-plane is
\begin{equation}\label{Helmholtz}
\left(\partial_{x}^2+\partial_{z}^2+k^2\right)
h=0,
\end{equation}
where for $p$-wave $h=h_y$, $k^2=\eps k_0^2$, $\eps=1$ in the slit and  $\eps_M$ in the metal. We introduce the polar coordinates with the center at intersection point of the walls extensions, Fig.~\ref{f:polar}. The Helmholtz equation (\ref{Helmholtz}) in free space $\varphi^2<\alpha^2$  reduces to
\begin{equation}\label{Helmholtz-polar}
\left[\frac1r\frac{\partial}{\partial r}\left(r\frac{\partial}{\partial r}\right)+
\frac1{r^2}\frac{\partial^2}{\partial\varphi^2}+k_0^2\right]H=0,
\end{equation}
where $k_0=\omega/c$.
We look for the solution of the form
$h(r,\varphi)=R(r)\Phi(\varphi)$ and get from (\ref{Helmholtz-polar}):
\begin{equation}\label{separation}
\frac{(rR')'}{rR}+\frac1{r^2}\frac{\Phi''}{\Phi}+k_0^2=0.
\end{equation}

The separation of variables occurs; the angular equation has form $\Phi''+\mu\Phi=0$, where $\mu$ is the separation parameter. At the boundaries of perfect conductor $(\ln\Phi)'(\varphi=\pm\alpha)=0.$  The even and odd modes with $\mu=-\pi^2m^2/\alpha^2$ are:
\begin{equation}\label{ideal-modes}
\Phi_m(\varphi)=\begin{cases}
\cos\frac{\pi m\varphi}{2\alpha},&m=0,2,\dots,\\
\sin\frac{\pi m\varphi}{2\alpha},&m=1,3,\dots
\end{cases}
\end{equation}
The general solution is a Fourier series, the decomposion over angular eigen functions
$\Phi_m(\varphi)$. Zero mode $m=0$ is a constant $\Phi_0=1$. The norm of solution is determined by the magnetic field $h_0$ at the entrance, like in unperturbed Eq. (\ref{unperturbed}). For small $\alpha$ we can approximately assume that the entrance is bounded from below by a horizontal line, and not by an arc.

For real metal the solution is more complicated, since the boundary conditions are different, namely, continuity of weighted normal derivative. Denoting separation parameter $\mu=-p^2$ in the slit and $\mu=-p_M^2$ in the metal we take into account the zero mode only for the sub-wavelength slit. Angular function for the zero mode at $\varphi>0$ is
\begin{equation}\label{Zero-mode-real}
\Phi_0(\varphi)=\begin{cases}
\cosh\left(p\frac{\varphi}{\alpha}\right),&\varphi<\alpha,\\
\exp\left(-p_M\frac{\varphi-\alpha}{\alpha}\right),&\alpha<\varphi.
\end{cases}
\end{equation}
At $\varphi<0$ the solution can be continued as an even function.
Then the radial equations in free space and metal are
\begin{equation}\label{radial}
R''+\frac1rR'+\left(k_0^2+\frac{p^2}{\alpha^2r^2}\right)R=0,\quad
R''+\frac1rR'+\left(k_M^2+\frac{p^2_M}{\alpha^2r^2}\right)R=0,
\end{equation}
respectively. In order to provide the solution as a cylindrical wave at infinity one must chose the Hankel function of the second kind:
\begin{equation}\label{Hankel}
R(r)=\begin{cases}
H^{(1)}_{ip/\alpha}(k_0r),&\varphi<\alpha,\\
H^{(1)}_{ip_M/\alpha}(k_Mr),&\alpha<\varphi.
\end{cases}
\end{equation}
We find the first relation between $p$ and $p_M$  equating the weighted logarithmic derivatives of angular function (\ref{Zero-mode-real}):
\begin{equation}\label{Border}
p\tanh p=-\frac{p_M}{\eps_M}.
\end{equation}

The second relation follows from radial solution (\ref{Hankel}). Since the entrance at $\alpha\ll1$ is far from the polar coordinate origin $r_0=\ell_0/\alpha$, where $\ell_0$ is the slit half-width, we have to use asymptotic formula for Hankel function (\ref{Hankel}). 
There are different asymptotic expansions of cylindrical functions $H^{(1)}_\nu(\xi)$. The choice of asymptotics is dictated by the relative rate of increasing in argument $\xi$ and parameter (the order) $\nu$. We need the Debye expansion \cite{W,BE2} while both the argument and parameter tend to infinity at fixed ratio $\nu/\xi$: 
\begin{equation}\label{Debye}
H_\nu^{(1)}\left(\frac{\nu}{\cos\psi}\right)\approx
\sqrt{\frac{2}{\pi\nu\tan\psi}}
\exp\left[i\nu\tan\psi-i\nu\psi-i\pi/4\right],\quad\nu\to\infty.
\end{equation}

Changing the variable $r=\ell_0/\alpha+z$ we write the argument of Hankel function in asymptotic formula  (\ref{Debye}) as
\begin{equation}\label{arg}
k_0\left(\frac{\ell_0}{\alpha}+z\right)
=\frac{\nu}{\cos\psi}.
\end{equation}
We expand $\psi$ to the Taylor series over the powers of $z$ 
limited ourselves by the main terms 
provided $z\ll r_0$ or
\begin{equation}\label{slow}
	\alpha z\ll\ell_0.
\end{equation}
Factor at the exponent (\ref{Debye}) in main order does not depend on $z$ and is included in an arbitrary wave amplitude $A$. The zeroth order in the exponent is also included in the amplitude. In the first approximation, the wave is described by the expression:
\begin{equation}
H_\nu^{(1)}\left(\frac{\nu}{\cos\psi}\right)\approx
Ae^{i\eta\nu z},\quad \eta=\left.\frac{d\left(\tan\psi-\psi\right)}{dz}
\right|_{z=0}.
\end{equation}
Differentiating expression (\ref{arg}), we get
\begin{equation}
H_\nu^{(1)}\left(\frac{\nu}{\cos\psi}\right)\approx
A\exp\left(\pm i\beta z \right),\quad
\beta = \frac{1}{\ell}\sqrt{a^2+p^2}.
\end{equation}
Here $a=k_0\ell<1$ is the dimensionless small parameter.
Analogously, for metal we have $\beta=\sqrt{\varepsilon a^2+p_M^2}/\ell$, where $\ell(z)=\ell_0+\alpha z$ is the slit half-width at height $z$, hence 
\begin{equation}\label{phase-velocity}
 p^2+a^2=p_M^2+\eps_Ma^2.
 \end{equation}
This is the second relation that means the equal propagation constants along $z$ in metal and free space 
or equal phase velocities. Substituting $p_M$ from (\ref{phase-velocity}) to (\ref{Border}) we obtain the dispersion relation
\begin{equation}\label{dispersion}
p\tanh p=-\frac{\sqrt{p^2+a^2(1-\eps_M)}}{\eps_M}.
\end{equation}
The similar relation has been derived for coupled surface plasmons at the parallel boundaries of metal plate surrounded by free space \cite{ZAYATS2005131,0034-4885-70-1-R01}.

The dispersion relation turns to be the same as for plane-parallel plates. The only difference consists in $z$-dependence of the half
width $\ell(z)$. Denote $L_z$ the height of the plates in $z$-direction. The variation of width up to $L_z$ should be small (\ref{slow}), i.e.
\begin{equation}\label{full-variation}
\alpha L_z\ll \ell_0.
\end{equation}
For $\alpha\neq0$ the Maxwell's tension  (\ref{sigma0}) can be integrated  over $z$:
\begin{eqnarray}
	f=2f_0\int\limits_0^{L_z}\frac{\exp\left(
		-\frac{-2\beta_2z}{1+\alpha z/\ell_0}\right)}{1+\alpha z/\ell_0}\beta_2\,dz\nonumber\\=
	\frac{2f_0\beta_2e^{-2\beta_2\ell_0/\alpha}}{\alpha}
	\left[
	\mbox{Ei}\left(\frac{-2\beta_2\ell_0}{\alpha}\right)-
	\mbox{Ei}\left(\frac{-2\beta_2\ell_0}{\alpha+\alpha^2L_z/\ell_0}\right)
	\right],\label{force}
\end{eqnarray}
where $f_0$ is the unperturbed force (\ref{unperturbed}), $\mbox{Ei}$ and is the integral exponent \cite{Olver10}:
\[
\mbox{Ei}\left(\zeta\right)=\int_{-\infty}^{\zeta}\frac{e^t}{t}dt.
\]

\begin{figure}
\includegraphics[width=0.6\textwidth]{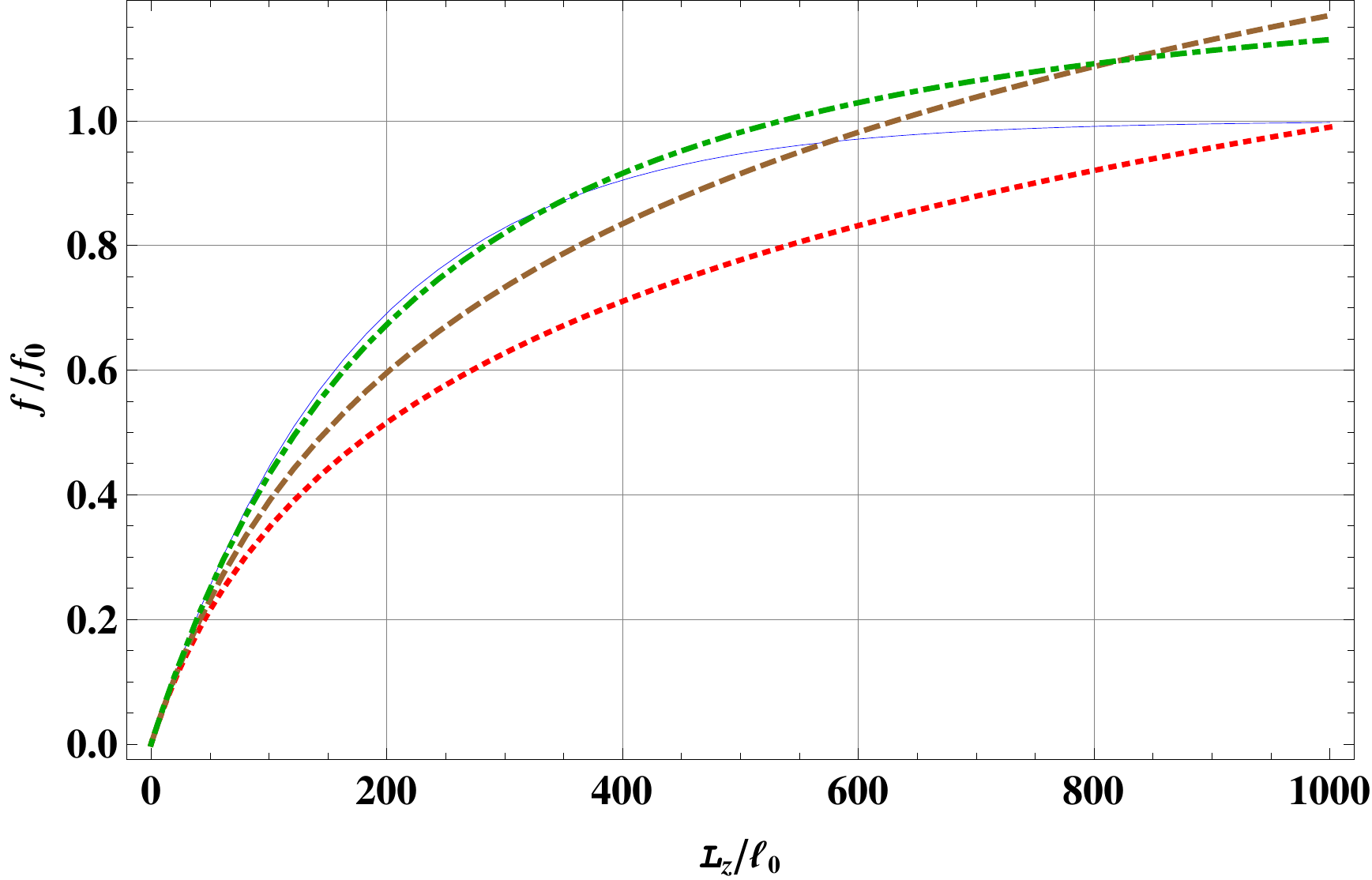}
\caption{The force $f/f_0$ for $\eps_M=-91.5+10.3i$ (gold at $\lambda=1.512~\mu$m \cite{Palik98} as a function of dimensionless height of plates $L_z/\ell_0$ at: $\alpha=0$ (solid line), $10^{-3}$ (dot-dash), $5\times10^{-3}$ (dashed), $10^{-2}$ (dotted).}\label{f:longitudinal} 	
\end{figure}

For small angle $\alpha\ll1$ the arguments of integral exponents become large parameters. The asymptotic of integral exponent gives a series for force (\ref{force}): 
\begin{equation}
	f=f_0\left(1-e^{-2\beta_2L_z}\right)\left\{1+\alpha
\left[(2\beta_2L_z)^{-1}+1+(2\beta_2L_z)\right]
\right\}+O(\alpha^2).
\end{equation}
A factor in front of curly bracket is a force between finite
parallel walls, the next term is linear correction for inclination:
\begin{equation}\label{correction}
	\delta f=-\frac{\alpha f_0}{2\ell_0}\beta_2L_z^2.
\end{equation}
The correction is negative, that means the decreasing force in broadened slit. Fig.~\ref{f:longitudinal} shows the dependence on  $L_z$. With decreasing angle $\alpha$ the curves tend to unperturbed dependence. In the domain of short slit the curves lie below the solid line, that corresponds to the negative correction (\ref{correction}). For tall slit, where the curves intersect solid line, our Taylor series obtained for small $z$ is not valid, since condition (\ref{full-variation}) being violated. 

\subsection{Transverse wedge}

We consider V-slit with wedge perpendicular to wave vector, Fig.~\ref{fig:longitudinal}~(b). Let us introduce the new variables at $y>0$
\[
x'=\frac{\ell x}{\ell+\alpha y},\quad y'=y
\]
to make the plates parallel in new coordinates.
The first-derivative operators transform to
\[
\partial_x=\frac{\ell}{\ell+\alpha y}\partial_{x'},\quad
\partial_y=-\frac{\alpha\ell x}{(\ell+\alpha y)^2}\partial_{x'}+\partial_{y'}.
\]
Hereafter the half-width $\ell\equiv\ell_0$ is independent of $z$.
Then the Laplace operator in old coordinates
acquires crossed terms in non-orthogonal varables:
\begin{equation}\label{skew}
\bigtriangleup=\frac{\ell^2}{(\ell+\alpha y)^2}\left[
1+\frac{\alpha^2 x^2}{(\ell+\alpha y)^2}\right]\partial_{x'}^2
+\partial_{y'}^2-\frac{2\alpha\ell x}{(\ell+\alpha y)^2}
\partial_{x'}\partial_{y'}+\frac{2\alpha^2\ell x}{(\ell+\alpha y)^3}\partial_{x'}+\partial_z^2.
\end{equation}

For $\alpha\ll1$ the Migdal's perturbation theory \cite{jM59} can be applied for geometry slightly different 
from exactly solvable (see also \cite[\S38]{LL3E}).
We seek the solution to (\ref{skew}) as a series:
\begin{equation}\label{series}
	h(x',y',z)=g_0+\alpha g_1+\alpha^2g_2+\dots
\end{equation}
Substituting (\ref{series}) into (\ref{skew}) and equating terms with the same power of $\alpha$ we get the chain.  Zero order yields the unperturbed equation $(\bigtriangleup+k^2)g_0=0$ with the known solution
\begin{equation}
	g_0=h_0e^{i\beta z}
	\begin{cases}
\cosh\varkappa x',&0<x'<l,\\
\cosh\varkappa le^{-\varkappa_M(x'-l)},&l\leqslant x'.
	\end{cases}
\end{equation}
The first order is $(\bigtriangleup+k^2)g_1\propto\partial_{x'}\partial_{y'}g_0=0$, since the zero-order solution is independent of $y'$, then $g_1=0$.  Correction $g_2$ has the second order in $\alpha$ and may be neglected. We do not present the second-order term here since it is negligibly small at $\alpha\ll1$.

\section{Off-normal incidence}\label{sec:oblique}

As it was shown in the previous studies \cite{0295-5075-94-6-64002,0957-0233-22-9-094008}, excitation of the zero mode in the narrow slit ($k_0 l < 1$) leads to attraction between the plates. Since in this case electro-magnetic pressure is proportional to $|\beta^{(0)}|^2 - |k_0|^2$ (where $\beta^{(0)}$ is a propagation constant of this mode), the fact that $|\beta^{(0)}|>k_0$ plays a crucial role. But for the others modes $|\beta^{(i)}|<k_0$, and then they are evanescent. Deviation of angle $\gamma$ between $\vec{k}_0$ and surface normal increases the contribution of odd modes, and then  decreases the attraction. We aim to calculate the effect of off-normal incidence below taking into account the first antisymmetric mode.

\subsection{Field in the slit}
Here we consider an incident $p$-wave, Fig.~\ref{fig:scheme}. The incident plane wave has only $y$-component of the magnetic field, and the wave vector $\vec{k_0} = (k_x, 0, k_z)$. The total field is $H(x,z,t) = H(x,z)e^{i\omega t}+\textrm{c.c.}$ Solution of the problem is based on  the two-dimensional Helmholtz equation (\ref{Helmholtz}) with the boundary conditions: continuity of tangential components of electric and magnetic fields at the boundaries.

We imply Fourier-transformation to the field in order to get algebraic equations instead of differential. Thus, the field in the slit transforms into Fourier series
\begin{equation}\label{eq:inslit}
H^{>} = \sum_{\nu} h_{\nu} b_{\nu}(x) e^{i\beta^{(\nu)} z},
\end{equation}
where $b_{\nu}(x) = \cos{q_{\nu}x} \ (\sin{q_{\nu}x})$ for the even (odd) modes, $\beta^{(\nu)2} = k_0^2-q_{\nu}^2$.
The field in free space transforms into Fourier integral:
\begin{equation}\label{eq:inspace}
H^{<} = (e^{i k_{0z} z}+R\ e^{-i k_{0z} z})\ e^{i k_{0x} x} + \int a_k e^{i k x - i \kappa z}\ dk,
\end{equation}
\[
R=\frac{\eps\cos\gamma-\sqrt{\eps-\sin^2\gamma}}{\eps\cos\gamma+\sqrt{\eps-\sin^2\gamma}}
\]
is the Fresnel reflection coefficient for $\ell\to0$, $\kappa^2 = k_0^2-k^2$.

The boundary conditions in the slit reveals dispersion equations for even modes:
\begin{equation}\label{eq:disp}
\tan{(q_{\nu}l)} = \frac{\sqrt{(1-\varepsilon)a^2 - (q_{\nu}\ell)^2}}{\varepsilon \ q_{\nu}\ell},
\end{equation}
for odd modes the trigonometric function has to be replaced: $\tan \to -\cot$.

As it was shown for perfect conductor, in the case of sub-wavelength slit and normal incidence it is sufficient to consider only the first mode, and the others are negligible \cite{sturman2010transmission,Shapiro16}. We imply the same assumption to the odd modes to simplify the calculations. We consider only the first even ($h_0$) and odd ($h_1$) modes:
\begin{equation}\label{2modes}
H^{>} = h_{0} \cosh{(q_0 x)} e^{i\beta^{(0)} z} + h_1\sin{(q_1 x)}e^{i\beta^{(1)} z},
\end{equation}
where $\beta^{(0)},\beta^{(1)}$ are the propagation constants of zero and first modes, respectively.
To satisfy the conditions at $z=0$, we follow the procedure  \cite{sturman2010transmission} $E_{x}^{<}(x) = E_{x}^{>}(x)$ expressing $a_k$ in terms of $h_0\ \text{and} \ h_1$. Then the continuity condition $H^{<}(x)=H^{>}(x), |x|<l$ gives coefficients
\begin{widetext}
\begin{eqnarray}
h_0 = \frac{(1+R)f_{q_0,k_{0x}} + \ell k_{0z}(1-R)/\pi \int \frac{\sinc((k_{0x}-k)\ell) f_{q_0,k} }{\kappa} dk }{f_{q_0,q_0} + \beta^{(0)} \ell/(2\pi)\left \{
\int \frac{f_{q_0,k} f_{q_0,k}}{\kappa} dk + \cosh{(q_0 \ell)}/\varepsilon \int \frac{G^{ev}(q_{0M,k}) f_{q_0,k}}{\kappa} dk  \right\}},\\
h_1 = \frac{(1+R)f_{q_1,-k_{0x}} + \ell k_{0z}(1-R)/\pi \int \frac{\sinc((k_{0x}-k)\ell) f_{q_1,-k} }{\kappa} dk }{i f_{q_1,q_1} + \beta^{(1)} \ell/(2\pi) \left\{
\int \frac{f_{q_1,k} f_{q_1,-k}}{\kappa} dk + \sin{(q_1 \ell)}/\varepsilon \int \frac{G^{odd}(q_{1M,k}) f_{q_1,-k}}{\kappa} dk  \right\}}.\label{eq:coef}
\end{eqnarray}
\end{widetext}
\begin{eqnarray*}
f_{q_0,k} = \sinc\left((i q_0+k)\ell\right) + \sinc\left((iq_0 - k)\ell\right),\\
f_{q_1,k} = i(\sinc\left((q_1 + k)\ell\right) - \sinc\left((q_1 - k)\ell)\right),\\
G^{ev}(x,y)= \frac{2(x\cos{y\ell} - 2y\sin{y\ell})}{\ell(x^2+y^2)},\\
G^{odd}(x,y)=\frac{-2i(x\sin{y\ell} + 2y\cos{y\ell})}{\ell(x^2+y^2)}
\end{eqnarray*}
and $q_{iM} =\sqrt{\beta^{(i)2}-\eps_M k_{0}^{2}}, i=0,1$ is the transverse wave vector in metal.

If $\gamma\ll 1$ it is possible to estimate the amplitudes. For perfect metal $q_0 \ell \to 0$, $q_1 \ell \to \pi/2$ then
\begin{equation}\label{eq:G2}
h_0 \approx 2-(k_0\ell \gamma)^2/3,~h_1 \approx  2 i  k_0\ell \gamma.
\end{equation}
Hence, the amplitude $h_1\sim \gamma$ while the correction to $h_0$ is of the order of $\gamma^2$.

\subsection{Maxwell tension tensor}

\begin{figure}
\includegraphics[width=0.6\textwidth]{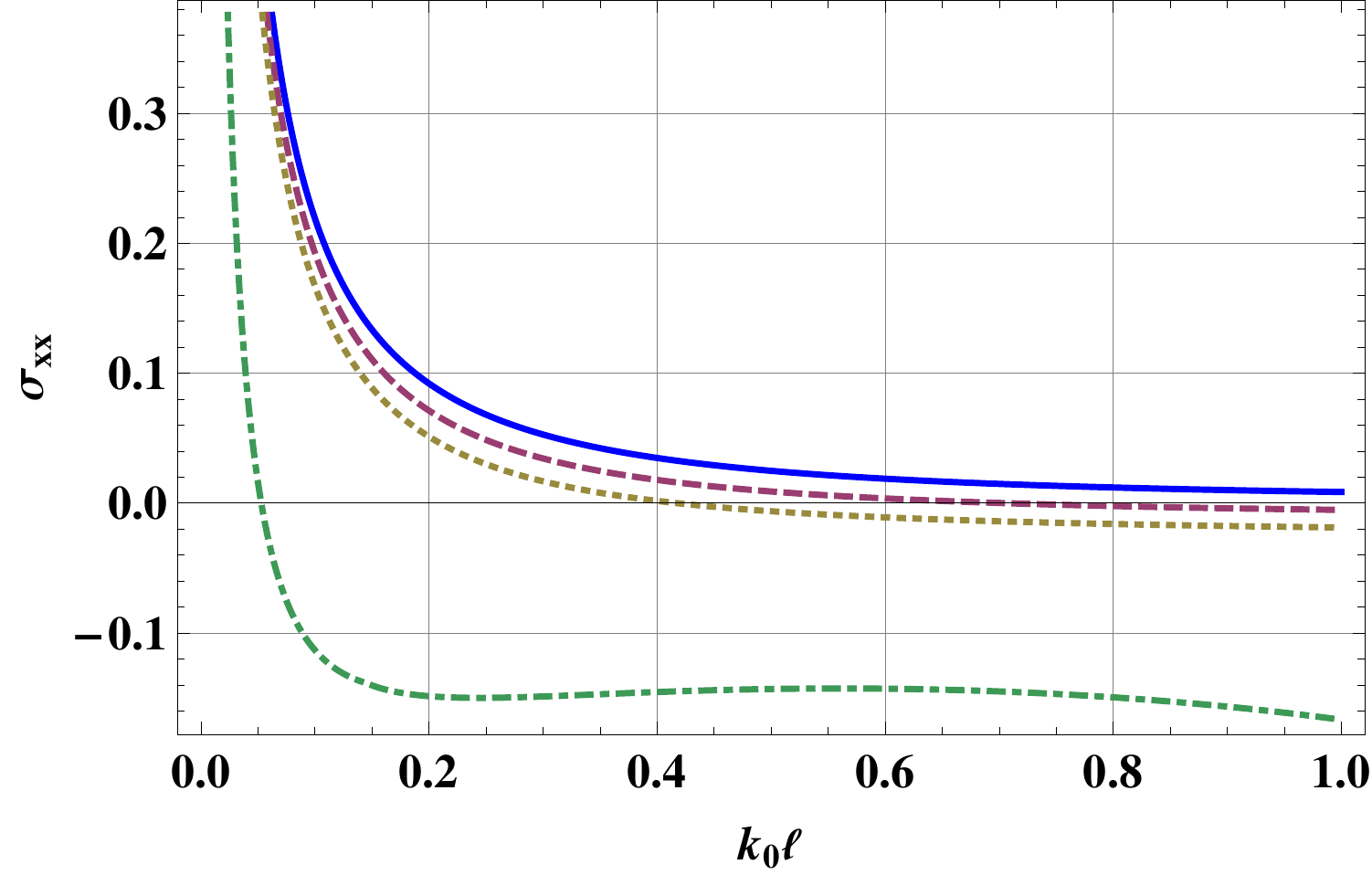}
	\caption{Dependence of $\sigma_{xx}(k_0 l)$ at  $\gamma=0$ (solid line),  $0.05$ (dashed), $0.1$ (doted), $1$ (dot-dashed).}
	\label{pic:sigma_al}
\end{figure}

Maxwell's tensor (\ref{sigma0}) determines attractive force between the walls. Non-zero $\gamma$ leads to excitation of the odd modes in the slit. But according to orthogonality of the even and odd modes, the set of equations on the Fourier coefficients splits into the two independent sets. We calculated the amplitudes of lower even and odd modes $h_0,h_1$. The presence of second mode adds the interference term $h_0^*h_1$ of the first order in $\gamma$.

We find the dependence of $\sigma_{xx}$ on $k_{0} \ell$ for different $\gamma$ numerically. Tension $\sigma_{xx}$ is shown in Fig.~\ref{pic:sigma_al}  for gold as a function of parameter $k_{0} \ell$. The first asymmetric mode decreases the force, especially for wider slit. Fig.~\ref{pic:sigma_al} demonstrates that the attraction can be changed by repulsion at higher width. When the curve intersects axis $x$ an equilibrium  distance appears being large for small angles and decreasing with $\gamma$.


\section{Conclusion}\label{sec:conclusions}

Three possible experimental uncertainties are considered: non-parallel plates along or transverse to wave propagation and off-normal incidence. We calculate their contribution to the force acting between the plates. Since in the first order force perturbation is a sum of all corrections, we calculate them separately. We get the analytic expression of field for wedge along the wave propagation. Correction to the force is of the first order in angle. For the transverse wedge  the correction is shown to be of the second order. Then it is necessary first to consider the contribution of longitudinal wedge. For non-parallel plates the sub-wavelength assumption $a=k_0\ell<1$ allows us to consider only the zero mode, but for the off-normal incidence we have to include the first odd mode. We calculate numerically the pressure acting on the walls.  Presence of the first odd mode decreases the pressure and could change attraction to repulsion.

\begin{acknowledgments}
We are grateful to V. Nesterov for excellent discussion of experimental parameters.
This work is supported by the Russian Foundation for Basic Research (\# 16-52-12026)
and The Council for grants of President of Russian Federation (NSh-6898.2016.2).
\end{acknowledgments}

\begin{thebibliography}{24}%
	\makeatletter
	\providecommand \@ifxundefined [1]{%
		\@ifx{#1\undefined}
	}%
	\providecommand \@ifnum [1]{%
		\ifnum #1\expandafter \@firstoftwo
		\else \expandafter \@secondoftwo
		\fi
	}%
	\providecommand \@ifx [1]{%
		\ifx #1\expandafter \@firstoftwo
		\else \expandafter \@secondoftwo
		\fi
	}%
	\providecommand \natexlab [1]{#1}%
	\providecommand \enquote  [1]{``#1''}%
	\providecommand \bibnamefont  [1]{#1}%
	\providecommand \bibfnamefont [1]{#1}%
	\providecommand \citenamefont [1]{#1}%
	\providecommand \href@noop [0]{\@secondoftwo}%
	\providecommand \href [0]{\begingroup \@sanitize@url \@href}%
	\providecommand \@href[1]{\@@startlink{#1}\@@href}%
	\providecommand \@@href[1]{\endgroup#1\@@endlink}%
	\providecommand \@sanitize@url [0]{\catcode `\\12\catcode `\$12\catcode
		`\&12\catcode `\#12\catcode `\^12\catcode `\_12\catcode `\%12\relax}%
	\providecommand \@@startlink[1]{}%
	\providecommand \@@endlink[0]{}%
	\providecommand \url  [0]{\begingroup\@sanitize@url \@url }%
	\providecommand \@url [1]{\endgroup\@href {#1}{\urlprefix }}%
	\providecommand \urlprefix  [0]{URL }%
	\providecommand \Eprint [0]{\href }%
	\providecommand \doibase [0]{http://dx.doi.org/}%
	\providecommand \selectlanguage [0]{\@gobble}%
	\providecommand \bibinfo  [0]{\@secondoftwo}%
	\providecommand \bibfield  [0]{\@secondoftwo}%
	\providecommand \translation [1]{[#1]}%
	\providecommand \BibitemOpen [0]{}%
	\providecommand \bibitemStop [0]{}%
	\providecommand \bibitemNoStop [0]{.\EOS\space}%
	\providecommand \EOS [0]{\spacefactor3000\relax}%
	\providecommand \BibitemShut  [1]{\csname bibitem#1\endcsname}%
	\let\auto@bib@innerbib\@empty
	\bibitem [{\citenamefont {Gardner}\ and\ \citenamefont
		{Varadan}(2001)}]{Gardner:2001}%
	\BibitemOpen
	\bibfield  {author} {\bibinfo {author} {\bibfnamefont {J.~W.}\ \bibnamefont
			{Gardner}}\ and\ \bibinfo {author} {\bibfnamefont {V.~K.}\ \bibnamefont
			{Varadan}},\ }\href@noop {} {\emph {\bibinfo {title} {Microsensors, Mems and
				Smart Devices}}}\ (\bibinfo  {publisher} {John Wiley \& Sons},\ \bibinfo
	{address} {New York},\ \bibinfo {year} {2001})\BibitemShut {NoStop}%
	\bibitem [{\citenamefont {Rebeiz}\ and\ \citenamefont
		{Muldavin}(2001)}]{969936}%
	\BibitemOpen
	\bibfield  {author} {\bibinfo {author} {\bibfnamefont {G.~M.}\ \bibnamefont
			{Rebeiz}}\ and\ \bibinfo {author} {\bibfnamefont {J.~B.}\ \bibnamefont
			{Muldavin}},\ }\href@noop {} {\bibfield  {journal} {\bibinfo  {journal} {IEEE
				Microwave Magazine}\ }\textbf {\bibinfo {volume} {2}},\ \bibinfo {pages} {59}
		(\bibinfo {year} {2001})}\BibitemShut {NoStop}%
	\bibitem [{\citenamefont {Rebeiz}(2004)}]{rebeiz2004rf}%
	\BibitemOpen
	\bibfield  {author} {\bibinfo {author} {\bibfnamefont {G.~M.}\ \bibnamefont
			{Rebeiz}},\ }\href@noop {} {\emph {\bibinfo {title} {RF MEMS: theory, design,
				and technology}}}\ (\bibinfo  {publisher} {John Wiley \& Sons},\ \bibinfo
	{address} {New York},\ \bibinfo {year} {2004})\BibitemShut {NoStop}%
	\bibitem [{\citenamefont {Ko}(2007)}]{ko2007trends}%
	\BibitemOpen
	\bibfield  {author} {\bibinfo {author} {\bibfnamefont {W.~H.}\ \bibnamefont
			{Ko}},\ }\href@noop {} {\bibfield  {journal} {\bibinfo  {journal} {Sensors
				and Actuators A: Physical}\ }\textbf {\bibinfo {volume} {136}},\ \bibinfo
		{pages} {62} (\bibinfo {year} {2007})}\BibitemShut {NoStop}%
	\bibitem [{\citenamefont {Grayson}\ \emph {et~al.}(2004)\citenamefont
		{Grayson}, \citenamefont {Shawgo}, \citenamefont {Johnson}, \citenamefont
		{Flynn}, \citenamefont {Li}, \citenamefont {Cima},\ and\ \citenamefont
		{Langer}}]{grayson2004biomems}%
	\BibitemOpen
	\bibfield  {author} {\bibinfo {author} {\bibfnamefont {A.~R.}\ \bibnamefont
			{Grayson}}, \bibinfo {author} {\bibfnamefont {R.~S.}\ \bibnamefont {Shawgo}},
		\bibinfo {author} {\bibfnamefont {A.~M.}\ \bibnamefont {Johnson}}, \bibinfo
		{author} {\bibfnamefont {N.~T.}\ \bibnamefont {Flynn}}, \bibinfo {author}
		{\bibfnamefont {Y.}~\bibnamefont {Li}}, \bibinfo {author} {\bibfnamefont
			{M.~J.}\ \bibnamefont {Cima}}, \ and\ \bibinfo {author} {\bibfnamefont
			{R.}~\bibnamefont {Langer}},\ }\href@noop {} {\bibfield  {journal} {\bibinfo
			{journal} {Proceedings of the IEEE}\ }\textbf {\bibinfo {volume} {92}},\
		\bibinfo {pages} {6} (\bibinfo {year} {2004})}\BibitemShut {NoStop}%
	\bibitem [{\citenamefont {Arias-Gonz\'{a}lez}\ and\ \citenamefont
		{Nieto-Vesperinas}(2003)}]{Arias-Gonzalez:03}%
	\BibitemOpen
	\bibfield  {author} {\bibinfo {author} {\bibfnamefont {J.~R.}\ \bibnamefont
			{Arias-Gonz\'{a}lez}}\ and\ \bibinfo {author} {\bibfnamefont
			{M.}~\bibnamefont {Nieto-Vesperinas}},\ }\href@noop {} {\bibfield  {journal}
		{\bibinfo  {journal} {J. Opt. Soc. Am. A}\ }\textbf {\bibinfo {volume}
			{20}},\ \bibinfo {pages} {1201} (\bibinfo {year} {2003})}\BibitemShut
	{NoStop}%
	\bibitem [{\citenamefont {Volpe}\ \emph {et~al.}(2006)\citenamefont {Volpe},
		\citenamefont {Quidant}, \citenamefont {Badenes},\ and\ \citenamefont
		{Petrov}}]{volpe2006surface}%
	\BibitemOpen
	\bibfield  {author} {\bibinfo {author} {\bibfnamefont {G.}~\bibnamefont
			{Volpe}}, \bibinfo {author} {\bibfnamefont {R.}~\bibnamefont {Quidant}},
		\bibinfo {author} {\bibfnamefont {G.}~\bibnamefont {Badenes}}, \ and\
		\bibinfo {author} {\bibfnamefont {D.}~\bibnamefont {Petrov}},\ }\href@noop {}
	{\bibfield  {journal} {\bibinfo  {journal} {Physical Review Letters}\
		}\textbf {\bibinfo {volume} {96}},\ \bibinfo {pages} {238101} (\bibinfo
		{year} {2006})}\BibitemShut {NoStop}%
	\bibitem [{\citenamefont {Chu}\ and\ \citenamefont
		{Mills}(2007)}]{chu2007laser}%
	\BibitemOpen
	\bibfield  {author} {\bibinfo {author} {\bibfnamefont {P.}~\bibnamefont
			{Chu}}\ and\ \bibinfo {author} {\bibfnamefont {D.}~\bibnamefont {Mills}},\
	}\href@noop {} {\bibfield  {journal} {\bibinfo  {journal} {Physical Review
				Letters}\ }\textbf {\bibinfo {volume} {99}},\ \bibinfo {pages} {127401}
		(\bibinfo {year} {2007})}\BibitemShut {NoStop}%
	\bibitem [{\citenamefont {Juan}\ \emph {et~al.}(2011)\citenamefont {Juan},
		\citenamefont {Righini},\ and\ \citenamefont {Quidant}}]{juan2011plasmon}%
	\BibitemOpen
	\bibfield  {author} {\bibinfo {author} {\bibfnamefont {M.~L.}\ \bibnamefont
			{Juan}}, \bibinfo {author} {\bibfnamefont {M.}~\bibnamefont {Righini}}, \
		and\ \bibinfo {author} {\bibfnamefont {R.}~\bibnamefont {Quidant}},\
	}\href@noop {} {\bibfield  {journal} {\bibinfo  {journal} {Nature Photonics}\
		}\textbf {\bibinfo {volume} {5}},\ \bibinfo {pages} {349} (\bibinfo {year}
		{2011})}\BibitemShut {NoStop}%
	\bibitem [{\citenamefont {Jones}\ \emph {et~al.}(2015)\citenamefont {Jones},
		\citenamefont {Marag{\`o}},\ and\ \citenamefont {Volpe}}]{jones2015optical}%
	\BibitemOpen
	\bibfield  {author} {\bibinfo {author} {\bibfnamefont {P.~H.}\ \bibnamefont
			{Jones}}, \bibinfo {author} {\bibfnamefont {O.~M.}\ \bibnamefont
			{Marag{\`o}}}, \ and\ \bibinfo {author} {\bibfnamefont {G.}~\bibnamefont
			{Volpe}},\ }\href@noop {} {\emph {\bibinfo {title} {Optical tweezers:
				Principles and applications}}}\ (\bibinfo  {publisher} {Cambridge University
		Press},\ \bibinfo {year} {2015})\BibitemShut {NoStop}%
	\bibitem [{\citenamefont {Nesterov}\ \emph {et~al.}(2011)\citenamefont
		{Nesterov}, \citenamefont {Frumin},\ and\ \citenamefont
		{Podivilov}}]{0295-5075-94-6-64002}%
	\BibitemOpen
	\bibfield  {author} {\bibinfo {author} {\bibfnamefont {V.}~\bibnamefont
			{Nesterov}}, \bibinfo {author} {\bibfnamefont {L.}~\bibnamefont {Frumin}}, \
		and\ \bibinfo {author} {\bibfnamefont {E.}~\bibnamefont {Podivilov}},\
	}\href@noop {} {\bibfield  {journal} {\bibinfo  {journal} {EPL (Europhysics
				Letters)}\ }\textbf {\bibinfo {volume} {94}},\ \bibinfo {pages} {64002}
		(\bibinfo {year} {2011})}\BibitemShut {NoStop}%
	\bibitem [{\citenamefont {Nesterov}\ and\ \citenamefont
		{Frumin}(2011)}]{0957-0233-22-9-094008}%
	\BibitemOpen
	\bibfield  {author} {\bibinfo {author} {\bibfnamefont {V.}~\bibnamefont
			{Nesterov}}\ and\ \bibinfo {author} {\bibfnamefont {L.}~\bibnamefont
			{Frumin}},\ }\href@noop {} {\bibfield  {journal} {\bibinfo  {journal}
			{Measurement Science and Technology}\ }\textbf {\bibinfo {volume} {22}},\
		\bibinfo {pages} {094008} (\bibinfo {year} {2011})}\BibitemShut {NoStop}%
	\bibitem [{\citenamefont {Nies}\ \emph {et~al.}(2016)\citenamefont {Nies},
		\citenamefont {Buetefisch}, \citenamefont {Naparty}, \citenamefont {Wurm},
		\citenamefont {Belai}, \citenamefont {Shapiro},\ and\ \citenamefont
		{Nesterov}}]{nies2016experimental}%
	\BibitemOpen
	\bibfield  {author} {\bibinfo {author} {\bibfnamefont {D.}~\bibnamefont
			{Nies}}, \bibinfo {author} {\bibfnamefont {S.}~\bibnamefont {Buetefisch}},
		\bibinfo {author} {\bibfnamefont {D.}~\bibnamefont {Naparty}}, \bibinfo
		{author} {\bibfnamefont {M.}~\bibnamefont {Wurm}}, \bibinfo {author}
		{\bibfnamefont {O.}~\bibnamefont {Belai}}, \bibinfo {author} {\bibfnamefont
			{D.}~\bibnamefont {Shapiro}}, \ and\ \bibinfo {author} {\bibfnamefont
			{V.}~\bibnamefont {Nesterov}},\ }in\ \href@noop {} {\emph {\bibinfo
			{booktitle} {SPIE Nanoscience+ Engineering}}}\ (\bibinfo {organization}
	{International Society for Optics and Photonics},\ \bibinfo {year} {2016})\
	p.\ \bibinfo {pages} {99222L}\BibitemShut {NoStop}%
	\bibitem [{\citenamefont {Nesterov}\ \emph {et~al.}(2016)\citenamefont
		{Nesterov}, \citenamefont {Belai}, \citenamefont {Nies}, \citenamefont
		{Buetefisch}, \citenamefont {Mueller}, \citenamefont {Ahbe}, \citenamefont
		{Naparty}, \citenamefont {Popadic},\ and\ \citenamefont
		{Wolff}}]{0026-1394-53-4-1031}%
	\BibitemOpen
	\bibfield  {author} {\bibinfo {author} {\bibfnamefont {V.}~\bibnamefont
			{Nesterov}}, \bibinfo {author} {\bibfnamefont {O.}~\bibnamefont {Belai}},
		\bibinfo {author} {\bibfnamefont {D.}~\bibnamefont {Nies}}, \bibinfo {author}
		{\bibfnamefont {S.}~\bibnamefont {Buetefisch}}, \bibinfo {author}
		{\bibfnamefont {M.}~\bibnamefont {Mueller}}, \bibinfo {author} {\bibfnamefont
			{T.}~\bibnamefont {Ahbe}}, \bibinfo {author} {\bibfnamefont {D.}~\bibnamefont
			{Naparty}}, \bibinfo {author} {\bibfnamefont {R.}~\bibnamefont {Popadic}}, \
		and\ \bibinfo {author} {\bibfnamefont {H.}~\bibnamefont {Wolff}},\
	}\href@noop {} {\bibfield  {journal} {\bibinfo  {journal} {Metrologia}\
		}\textbf {\bibinfo {volume} {53}},\ \bibinfo {pages} {1031} (\bibinfo {year}
		{2016})}\BibitemShut {NoStop}%
	\bibitem [{\citenamefont {Shapiro}\ \emph {et~al.}(2016)\citenamefont
		{Shapiro}, \citenamefont {Nies}, \citenamefont {Belai}, \citenamefont
		{Wurm},\ and\ \citenamefont {Nesterov}}]{Shapiro16}%
	\BibitemOpen
	\bibfield  {author} {\bibinfo {author} {\bibfnamefont {D.}~\bibnamefont
			{Shapiro}}, \bibinfo {author} {\bibfnamefont {D.}~\bibnamefont {Nies}},
		\bibinfo {author} {\bibfnamefont {O.}~\bibnamefont {Belai}}, \bibinfo
		{author} {\bibfnamefont {M.}~\bibnamefont {Wurm}}, \ and\ \bibinfo {author}
		{\bibfnamefont {V.}~\bibnamefont {Nesterov}},\ }\href@noop {} {\bibfield
		{journal} {\bibinfo  {journal} {Opt. Express}\ }\textbf {\bibinfo {volume}
			{24}},\ \bibinfo {pages} {15972} (\bibinfo {year} {2016})}\BibitemShut
	{NoStop}%
	\bibitem [{\citenamefont {Watson}(1952)}]{W}%
	\BibitemOpen
	\bibfield  {author} {\bibinfo {author} {\bibfnamefont {G.~N.}\ \bibnamefont
			{Watson}},\ }\href@noop {} {\emph {\bibinfo {title} {A treatise on the theory
				of Bessel functions}}}\ (\bibinfo  {publisher} {The University press},\
	\bibinfo {address} {Cambridge},\ \bibinfo {year} {1952})\BibitemShut
	{NoStop}%
	\bibitem [{\citenamefont {Bateman}\ and\ \citenamefont {Erdelyi}(1953)}]{BE2}%
	\BibitemOpen
	\bibfield  {author} {\bibinfo {author} {\bibfnamefont {H.}~\bibnamefont
			{Bateman}}\ and\ \bibinfo {author} {\bibfnamefont {A.}~\bibnamefont
			{Erdelyi}},\ }\href@noop {} {\emph {\bibinfo {title} {Higher transcendental
				functions}}},\ Vol.~\bibinfo {volume} {2}\ (\bibinfo  {publisher}
	{McGraw-Hill},\ \bibinfo {address} {New York},\ \bibinfo {year}
	{1953})\BibitemShut {NoStop}%
	\bibitem [{\citenamefont {Zayats}\ \emph {et~al.}(2005)\citenamefont {Zayats},
		\citenamefont {Smolyaninov},\ and\ \citenamefont
		{Maradudin}}]{ZAYATS2005131}%
	\BibitemOpen
	\bibfield  {author} {\bibinfo {author} {\bibfnamefont {A.~V.}\ \bibnamefont
			{Zayats}}, \bibinfo {author} {\bibfnamefont {I.~I.}\ \bibnamefont
			{Smolyaninov}}, \ and\ \bibinfo {author} {\bibfnamefont {A.~A.}\ \bibnamefont
			{Maradudin}},\ }\href@noop {} {\bibfield  {journal} {\bibinfo  {journal}
			{Physics Reports}\ }\textbf {\bibinfo {volume} {408}},\ \bibinfo {pages} {131
		} (\bibinfo {year} {2005})}\BibitemShut {NoStop}%
	\bibitem [{\citenamefont {Pitarke}\ \emph {et~al.}(2007)\citenamefont
		{Pitarke}, \citenamefont {Silkin}, \citenamefont {Chulkov},\ and\
		\citenamefont {Echenique}}]{0034-4885-70-1-R01}%
	\BibitemOpen
	\bibfield  {author} {\bibinfo {author} {\bibfnamefont {J.~M.}\ \bibnamefont
			{Pitarke}}, \bibinfo {author} {\bibfnamefont {V.~M.}\ \bibnamefont {Silkin}},
		\bibinfo {author} {\bibfnamefont {E.~V.}\ \bibnamefont {Chulkov}}, \ and\
		\bibinfo {author} {\bibfnamefont {P.~M.}\ \bibnamefont {Echenique}},\
	}\href@noop {} {\bibfield  {journal} {\bibinfo  {journal} {Reports on
				Progress in Physics}\ }\textbf {\bibinfo {volume} {70}},\ \bibinfo {pages}
		{1} (\bibinfo {year} {2007})}\BibitemShut {NoStop}%
	\bibitem [{\citenamefont {Ovler}\ \emph {et~al.}(2010)\citenamefont {Ovler},
		\citenamefont {Lozier}, \citenamefont {Boisvert},\ and\ \citenamefont
		{Clark}}]{Olver10}%
	\BibitemOpen
	\bibfield  {author} {\bibinfo {author} {\bibfnamefont {F.~W.~J.}\
			\bibnamefont {Ovler}}, \bibinfo {author} {\bibfnamefont {D.~W.}\ \bibnamefont
			{Lozier}}, \bibinfo {author} {\bibfnamefont {R.~F.}\ \bibnamefont
			{Boisvert}}, \ and\ \bibinfo {author} {\bibfnamefont {C.~W.}\ \bibnamefont
			{Clark}},\ }\href@noop {} {\emph {\bibinfo {title} {NIST Handbook of
				Mathematical Functions}}}\ (\bibinfo  {publisher} {Cambridge University
		Press},\ \bibinfo {address} {New York},\ \bibinfo {year} {2010})\BibitemShut
	{NoStop}%
	\bibitem [{\citenamefont {Palik}(1998)}]{Palik98}%
	\BibitemOpen
	\bibinfo {editor} {\bibfnamefont {E.~D.}\ \bibnamefont {Palik}},\ ed.,\
	\href@noop {} {\emph {\bibinfo {title} {Handbook of Optical Constants of
				Solids. Vol 1,2.}}}\ (\bibinfo  {publisher} {Academic Press},\ \bibinfo
	{address} {London},\ \bibinfo {year} {1998})\BibitemShut {NoStop}%
	\bibitem [{\citenamefont {Migdal}(1960)}]{jM59}%
	\BibitemOpen
	\bibfield  {author} {\bibinfo {author} {\bibfnamefont {A.~B.}\ \bibnamefont
			{Migdal}},\ }\href@noop {} {\bibfield  {journal} {\bibinfo  {journal} {JETP}\
		}\textbf {\bibinfo {volume} {37}},\ \bibinfo {pages} {176} (\bibinfo {year}
		{1960})}\BibitemShut {NoStop}%
	\bibitem [{\citenamefont {Landau}\ and\ \citenamefont {Lifshitz}(1977)}]{LL3E}%
	\BibitemOpen
	\bibfield  {author} {\bibinfo {author} {\bibfnamefont {L.~D.}\ \bibnamefont
			{Landau}}\ and\ \bibinfo {author} {\bibfnamefont {E.~M.}\ \bibnamefont
			{Lifshitz}},\ }\href@noop {} {\emph {\bibinfo {title} {Quantum Mechanics
				(Non-relativistic Theory)}}},\ \bibinfo {series} {Course of Theoretical
		Physics}, Vol.~\bibinfo {volume} {3}\ (\bibinfo  {publisher} {Elsevier},\
	\bibinfo {address} {New York},\ \bibinfo {year} {1977})\BibitemShut {NoStop}%
	\bibitem [{\citenamefont {Sturman}\ \emph {et~al.}(2010)\citenamefont
		{Sturman}, \citenamefont {Podivilov},\ and\ \citenamefont
		{Gorkunov}}]{sturman2010transmission}%
	\BibitemOpen
	\bibfield  {author} {\bibinfo {author} {\bibfnamefont {B.}~\bibnamefont
			{Sturman}}, \bibinfo {author} {\bibfnamefont {E.}~\bibnamefont {Podivilov}},
		\ and\ \bibinfo {author} {\bibfnamefont {M.}~\bibnamefont {Gorkunov}},\
	}\href@noop {} {\bibfield  {journal} {\bibinfo  {journal} {Physical Review
				B}\ }\textbf {\bibinfo {volume} {82}},\ \bibinfo {pages} {115419} (\bibinfo
		{year} {2010})}\BibitemShut {NoStop}%
\end{thebibliography}
%

\end{document}